\documentclass[draftclsnofoot,onecolumn,12pt,twoside]{IEEEtran}
\usepackage{tcolorbox}
\usepackage[noadjust]{cite}
\usepackage[cmex10]{amsmath}
\usepackage{amssymb,amsfonts,amscd,amsthm,amsbsy}
\usepackage{graphicx}
\usepackage{stmaryrd}
\usepackage[mathscr]{eucal}
\usepackage{mathtools}
\usepackage{bbm}
\usepackage{xcolor}
\usepackage{algpseudocode}
\usepackage{enumerate}
\usepackage{hyperref}

\usepackage{amssymb, amsfonts, amscd, amsthm, amsbsy, amsmath}
\usepackage{subcaption}
\usepackage{array}
\usepackage{float}
\usepackage{algpseudocode}
\usepackage{color, colortbl}
\definecolor{Gray}{gray}{0.9}
\usepackage{multirow}
\usepackage{multicol}
\usepackage{booktabs,ragged2e}
\newtheorem{theorem}{Theorem}[section]
\newtheorem{lemma}{Lemma}[section]

\newtheorem{claim}{Claim}[subsection]
\newtheorem{remark}{Remark}[section]

\theoremstyle{definition}

\usepackage{etoolbox}
\AtEndEnvironment{example}{\null\hfill$\square$}%
\AtEndEnvironment{remark}{\null\hfill$\square$}%
\let\emptyset\varnothing
\newcommand{\Fb}{\mathbbmss{F}} % field
\newcommand{\lset}{\llbracket}  % [[
\newcommand{\rset}{\rrbracket}  % ]]
\newcommand{\wt}{w_H}   % weight
\newcommand{\RM}{\mathrm{RM}}   % Reed-Muller code
    
\newcommand{\p}{\pmb}           % bold font

\newcommand{\integerbox}[1]{\lset #1 \rset}

\newcommand{\supp}{\mathsf{supp}}

\newcommand{\rpir}{R_\mathsf{PIR}} %PIR Rate
\newcommand{\rst}{R_\mathsf{St}} %Storage rate
\newcommand{\dber}[3]{\mathsf{DB}_{#1}(#2, #3)} % Dual Berman Code, the first parameter is n, next is r, final is m. 
\newcommand{\ber}[3]{\mathsf{B}_{#1}(#2, #3)} % Berman Code, the first parameter is n, the next is r, and the final is m. 
\newcommand{\basisdber}[3]{B_{\mathsf{DB}_{#1}}(#2, #3)}
\newcommand{\basisber}[3]{B_{\mathsf{B}_{#1}}(#2, #3)}
\title{$t$-PIR Schemes with Flexible Parameters via Star Products of Berman Codes}
\author{Srikar Kale, Keshav Agarwal and Prasad Krishnan% <-this % stops a space
\thanks{\hrule}%

\thanks{Srikar, Keshav and Dr. Krishnan are with the Signal Processing and Communications Research Center, International Institute of Information Technology, Hyderabad, 500032, India (email: $\{$ srikar.kale@research., keshav.agarwal@student., prasad.krishnan@$\}$iiit.ac.in). %Acknowledgment:  Dr. Krishnan acknowledges support from SERB-DST project CRG/2019/005572.}
}
}

\begin{document}

\maketitle
\vspace{-0.3cm}

\begin{abstract}
We present a new class of private information retrieval (PIR) schemes that keep the identity of the file requested private in the presence of at most $t$ colluding servers, based on the recent framework developed for such $t$-PIR schemes using star products of transitive codes. These $t$-PIR schemes employ the class of Berman codes as the storage-retrieval code pairs. Berman codes, which are binary linear codes of length $n^m$ for any $n\geq 2$ and $m\geq 1$ being positive integers, were recently shown to achieve the capacity of the binary erasure channel. We provide a complete characterization of the star products of the Berman code pairs, enabling us to calculate the PIR rate of the star product-based schemes that employ these codes. The schemes we present have flexibility in the number of servers, the PIR rate, the storage rate, and the collusion parameter $t$, owing to numerous codes available in the class of Berman codes.

\end{abstract}

\section{Introduction} \label{sec: introduction}

The \textit{private information retrieval} (PIR) scenario, starting from the original work by Chor et al.\cite{PIR1, PIR2}, consists of one or more servers that hold a library (or a \textit{database}) of $M$ files and a client which wants to retrieve a file from this library. The privacy requirement is that the client does not want to reveal the identity of the file requested to any of the servers. For a specific storage technique employed at the servers, a PIR protocol is a design of privacy-preserving client queries, so that the requested file can be downloaded from the server's responses. Typically, the download communication cost from the server(s) to the client is much higher than the upload costs. Thus, the PIR rate is the ratio of the size of the file requested to the total number of bits downloaded from the servers. Thus, the goal of designing a good PIR protocol is to maximize the PIR rate without compromising the privacy requirement.

For the storage scenario when each server stores the entire library, that is, the files being replicated across the servers, the maximum rate of PIR (PIR \textit{capacity}) was characterized in \cite{SunJafarCapacity}. In \cite{SunJafarCapacity}, the authors derived a lower bound on the PIR capacity for such a replication-based database and presented a PIR scheme that remarkably achieves the lower bound using interference alignment techniques. Subsequent works studied the PIR problem for non-replicated storage systems, especially where the files may be stored in the servers using a linear code. PIR protocols for the MDS-coded storage were presented in  \cite{Tajeddine_MDSCOded_2018}. The MDS-PIR capacity, i.e., the PIR capacity for the MDS-coded storage system, was derived in \cite{Banawan_ulukus_PIRCoded_MDS_2018}. It was further shown that MDS-coded storage is not strictly necessary to achieve the capacity equal to the MDS-PIR capacity \cite{PIR_linearcodes_arbit}. While the above results focus on privacy in the \textit{non-colluding} scenario (that is, when the assumption is that servers do not communicate with each other), the capacity of PIR for replicated databases in the $t$-collusion scenario (where up to $t$ servers can \textit{collude}, i.e., exchange query data) was shown in \cite{SunJafar_Capacity_Colluding}. However, to the best of our knowledge, the PIR capacity for MDS-coded storage and $t$-collusion remains unknown in the general scenario (for recent progress on this question, see, for instance, \cite{Holzbaur_et_al_TowardsCodedColludingCapacity}). In recent years, the information-theoretic characterization of the PIR problem has been carried over to several new scenarios (see  \cite{ulukus_survey_2022_PIR} for a recent survey).

A general framework for employing linear codes for storage and retrieval in a PIR scenario with server collusion was presented in \cite{FriejHollantiSIAM}. In the scheme presented in \cite{FriejHollantiSIAM}, the authors consider that each file is encoded via a $(n,k)$ linear \textit{storage code} $\cal C$ and placed in $n$ servers. During the retrieval phase, the client constructs its queries by employing a \textit{retrieval code}, which is also a $n$-length linear code $\cal D$. The responses from each server are dot-products of the query vector sent to the server, with its stored data. For such a PIR protocol, it was shown that the colluding parameter $t$ (quantum of collusion that the protocol can ensure privacy against) depends on the minimum distance of the dual of the retrieval code.
Further, the PIR rate of this scheme can be characterized using the parameters of the so-called \textit{star product} (also called the Schur product) of the two codes, denoted by ${\cal C}\star {\cal D}$. While computing the star product of two arbitrary codes may be challenging in general, the case of storage-retrieval code pairs being Generalized Reed-Solomon (GRS) codes was presented in \cite{FriejHollantiSIAM}, for which the PIR rate was obtained and shown to be asymptotically optimal (as the number of files grows large) in some cases. The work \cite{tPIRIEEE} took this model forward for transitive codes in general, and the PIR parameters resulting from choosing Reed-Muller ($\RM$) codes (which are transitive) for the storage-retrieval code pairs were presented in \cite{tPIRIEEE}. This was further extended to PIR schemes which tolerate a finite number of non-responsive and Byzantine (i.e., erroneously responding) servers \cite{PIRRMIEEE}. Note that using these codes comes with some restrictions on some parameters of the PIR schemes possible. For instance, the $n$-length GRS codes require a field size that is $\Theta(n)$, while using $\RM$ codes requires the code length (and hence the number of servers) to be a power-of-$2$.   Recent work following this line of research has also looked at algebraic geometry codes, and cyclic codes \cite{HaoCHen22, ChenXu22} for use as storage-retrieval code pairs. 

In the current work, we present new PIR schemes using the framework of \cite{FriejHollantiSIAM}. These schemes employ the class of Berman codes for the storage-retrieval code pairs. Berman codes were recently identified in work \cite{DBLakPra}, in which they were shown to achieve the capacity of the binary erasure channel. These codes are generalizations of $\RM$ codes and of a class of Abelian codes constructed by Berman \cite{Ber_Cybernetics_II_67}. Berman codes are transitive binary linear codes of length $n^m$ (for any $n\geq 2$ and $m\geq 1$ being integers). In this work, we give a complete characterization of the star products of these codes. This characterization, coupled with the fact that these codes are transitive, enables us to employ them effectively in the star product PIR framework of \cite{FriejHollantiSIAM}, resulting in new PIR schemes. One key advantage of these new schemes is that the parameters $n$ and $m$ can be freely chosen, which gives us a large class of PIR schemes for a flexible number of servers, with various achievable triples of the storage code rate, PIR rates, and collusion parameter $t$. Further, as these codes are binary, the resultant PIR schemes require only a binary field, in contrast to the potential higher field size requirements of GRS codes-based schemes of \cite{FriejHollantiSIAM}. The $\RM$ codes-based PIR schemes from \cite{tPIRIEEE} are obtained as special cases of the Berman codes-based schemes we present in this work. 

Our work is organized as follows. In Section \ref{sec:  Preliminaries}, we first review the PIR schemes using star products of transitive codes as introduced in \cite{FriejHollantiSIAM,tPIRIEEE} (Subsection \ref{subsec: starprodPIRreview}). We then recall relevant definitions and results from \cite{DBLakPra} regarding Berman codes (Subsection \ref{subsec: bermancodesprelims}). In Section \ref{sec: starprodBermanCodes}, we derive the structure of the star products between various combinations of Berman codes. In Section \ref{sec: tPIRusingBerandDBer}, we present the resulting new PIR schemes that employ codes from the Berman family for the storage-retrieval code pairs and present numerical results comparing their performance with the $\RM$ codes-based schemes of \cite{tPIRIEEE}. We present some discussions for future work along these lines in Section \ref{sec: discussion} and conclude this work.

\emph{Notation: } We borrow most of the notation from \cite{DBLakPra}, as they are required to recall the definitions and relevant results of Berman codes. Let $\lset k \rset$ denote the set $\{0, 1, \dots, k-1\}$, and $[k]$ denote the set $\{1,\hdots,k\}$, for any positive integer $k$. The notation $\emptyset$ denotes the empty set. For sets, $A, B$ we define $A \setminus B$ as the set of elements in $A$ but not in $B$. For any vector $\p{x}$, $\supp ({\p x})$ denotes the support set of ${\p x}$. The zero-vector is denoted by $\p 0$. The concatenation of two vectors ${\p a}, {\p b}$ is denoted by, $({\p a}|{\p b}). $  The span of a collection of vectors $B$ from a vector space is denoted as $span(B)$. The dimension of a code ${\cal C}$ is denoted by $\dim({\cal C})$. The dual of a linear code $\cal C$ is denoted as ${\cal C}^\perp$. The Hamming weight of a vector ${\p a}$ is denoted by $\wt({\p a})$.  The minimum distance of a code ${\cal C}$ is denoted by, $d_{\min}({\cal C}). $ The binomial coefficient is denoted by $\binom{n}{k}$.   A $n$-length vector ${\p a}$ is identified on occasion by its components, as ${\p a}=(a_i: i\in \integerbox{n}). $  For some, $S\subset\integerbox{n}, $ we denote the sub-vector with components $a_i: i\in S$ as ${\p a}_S$. If ${\p a}$ is a $n^m$-length vector for some $m\geq 1$, we also use the concatenation representation ${\p a}=({\p a}_0|{\p a}_1|\hdots|{\p a}_{n-1})$, where ${\p a}_l: l\in\integerbox{n}$ are subvectors of length $n^{m-1}$. The binary field is denoted as $\Fb_2$, and an arbitrary field as $\Fb$. In some places, we use the abbreviation `s.t.' to denote the phrase `such that'.

\section{Preliminaries} 
\label{sec:  Preliminaries}
\subsection{$t$-PIR Schemes from Star Products :  Review of \cite{FriejHollantiSIAM,tPIRIEEE}}
\label{subsec: starprodPIRreview}
We now briefly review the PIR problem as presented in \cite{FriejHollantiSIAM} and the relevant results from \cite{FriejHollantiSIAM,tPIRIEEE} for our purposes.

We consider a coded data storage scheme in $n_s$ servers, using a linear code ${\cal C}\subset {\Fb}^{n_s}$ (called the \textit{storage code}) of dimension $k_{\cal C}$ and length $n_s$. We denote the rate of the storage code $\frac{k_{\cal C}}{n_s}$ as $\rst$. Consider $M$ files denoted as $X_i\in  \Fb^{b \times k_{\cal C}}: \forall i\in [M]$. The encoding and placement of the coded files are done as follows. Let $G_{\cal C}$ be a generator matrix of the storage code $\cal C$. The files are stacked as  a $Mb\times k_{\cal C}$ matrix $X=[X_1^T,\hdots,X_M^T]^T$. The encoding is then done by the matrix multiplication $Y=XG_{\cal C}$, where $Y$ is the resultant matrix $Mb\times n_s$. Then, the $i^{th}$ column of $Y$, denoted by $Y_i\in \Fb^{Mb\times 1}$, is stored in the $i^{th}$ server, for each $i\in[n_s]$. 

We now recount the retrieval process of a linear PIR scheme. The client wants to retrieve a file $X_d$ (where $d$ refers to the demanded file index) without revealing $d$ to some collections of subsets of the servers, to be discussed later. In order to do so, the client uses some private randomness to construct the query matrix $\p Q^{(d)}=\left(\p q_{1}^{(d)}, \ldots, \p q_{n_s}^{(d)}\right)$, where the query $\p q_{i}^{(d)}\in\Fb^{bM \times 1}$ will be sent to the server $i$, for each $i\in[n_s].$ The server $i$ responds to this query $\p q_{i}^{(d)}$ with the response $\p r_{i}^{(d)}=\langle\p q_{i}^{(d)},Y_i\rangle$, where $\langle~,~\rangle$ denotes the dot-product. This process of query-response can continue between the client and the servers for some $S$ iterations, such that the file $X_d$ can be reconstructed at the client (using some appropriate reconstruction function) from the $S\cdot n_s$ responses from the servers. A \textit{colluding} set of servers $T\subset[n_s]$ is a subset of servers that can communicate (the queries sent to each of them) with each other to deduce the requested file index $d$. We are interested in PIR schemes that keep the identity of $d$ private against  the collusion of every set of servers of size $t$. In other words, for each such colluding set $T$ in the collection $\{T\subset[n_s]\colon |T|=t\}$, the PIR scheme satisfies $I(\{\p q_{i}^{(d)}: \forall i\in T\}~ ;~ d)=0$, where $I(~;~)$ denotes the mutual information. Such PIR schemes are also referred to as $t$-PIR schemes.  The rate of the PIR scheme is the ratio of the size of the file requested to the total number of bits downloaded from the servers. For this above-described scheme, the rate  is then denoted as $\rpir\triangleq \dfrac{bk_{\cal C}}{Sn_s}$. Clearly, we are interested in high-rate PIR schemes. 

We now recollect briefly the role of star products in constructing $t$-PIR schemes, as presented in described in \cite{FriejHollantiSIAM,tPIRIEEE}. Let ${\cal D}\subset \Fb^{n_s}$ be a $k_{\cal D}$ dimensional code with length $n_s$. The star product of the codewords $\p c=(c_1,\hdots,c_{n_s})\in \cal C$ and $\p d=(d_1,\hdots,d_{n_s})\in \cal D$ is defined as $\p c\star \p d\triangleq (c_1 d_1, \hdots, c_{n_s} d_{n_s}).$  Observe that the star product is commutative, associative, and distributive across vector addition. The star product of the two codes $\cal C$ and $\cal D$ is defined as 
\begin{align*}
        {\cal C} \star {\cal D}  = span\{ {\p c} \star {\p d} \colon \forall \; {\p c} \in {\cal C}, \forall {\p d} \in {\cal D}\}.        
\end{align*}
Note that the star product between codes is also commutative and associative.

In the retrieval scheme of \cite{FriejHollantiSIAM,tPIRIEEE}, the client designs the query matrix (for each iteration) $\p Q^{(d)}\in \Fb^{Mb\times n_s}$ as a sum of two matrices, i.e., $\p Q^{(d)}=\p D+\p E^{(d)}$, where each row of the matrix $\p D\in \Fb^{Mb\times n_s}$ is generated uniformly at random from the code $\cal D$. The code $\cal D$ is known as the \textit{retrieval code}. The matrix $\p E^{(d)}$ is to be designed carefully in order to obtain some coded symbols of the desired file, which are useful for recovering the symbols of the desired file that have not been obtained in prior iterations. It is not difficult to see that the response vector will be a linear combination of the star products of the rows of $\p Q^{(d)}$ and the rows of the stored codeword matrix $Y$ (which are codewords of $\cal C$). Post download, the decoding step involves a process that is similar to syndrome decoding of the code ${\cal C}\star{\cal D}$ (for details regarding the decoding, we refer the reader to \cite{FriejHollantiSIAM,tPIRIEEE}). This scheme shown in \cite{FriejHollantiSIAM,tPIRIEEE} results in the following theorem (which we rephrase slightly for our purposes).

\begin{theorem}[\cite{tPIRIEEE}, Theorem 4]
\label{Theorem:  Starproduct}
Let $\cal C$ and $\cal D$ be linear codes in $\Fb^{n_s}$ such that the codes $\cal C$ and ${\cal C}\star {\cal D}$ are both transitive\footnote{We say that a code $\cal C$ is transitive if its automorphism group is transitive.}. Then, there is  a $t$-PIR scheme for the distributed storage system $Y=XG_{C}$ with a PIR rate $ \rpir = \frac{\dim((\cal C \star \cal D)^{\perp})}{n_s} $ which protects against $t =(d_{\min}({\cal D^{\perp}})-1)$ -collusion, which uses the code $\cal D$ as the retrieval code. 
\end{theorem}

\subsection{Berman Codes and their Dual Codes}
\label{subsec: bermancodesprelims}
We recall the definitions and relevant properties of codes identified in \cite{DBLakPra} termed as Berman and Dual Berman codes. These codes are a general class of binary linear codes which include the class of Abelian group codes constructed in \cite{Ber_Cybernetics_II_67, BlakNor}, using a group algebra framework, as well as the $\RM$ codes.  These results are from \cite{DBLakPra}, and we refer the interested reader to the same for more details. 
\subsubsection{{Recursive Definition of Berman \text{and} Dual Berman Codes}} \label{sec: sub: recursive_defn}
For some positive integers $n\geq 2$ and $m$, for some non-negative integer $r$ such that $0 \leq r\leq m$, the family of codes $\ber{n}{r}{m}\subseteq \Fb_2^{n^m}$, called the \emph{Berman codes with parameters $n, m$, and $r$}, is defined as
\begin{align*}
    &\ber{n}{m}{m} \triangleq \{{\p 0}\in \Fb_2^{n^m}\}, \\
    &\ber{n}{0}{m} \triangleq \{{\p c}\in \Fb_2^{n^m}\colon \sum_i c_i=0\},
\end{align*} and for $m\geq 2$ and $1\leq r\leq m-1, $
\begin{align}
\label{BermanRecDef}
    \ber{n}{r}{m}\triangleq \{&({ \p v}_0|{ \p v}_1|\hdots|{ \p v}_{n-1})\colon { \p v}_l\in \ber{n}{r-1}{m-1}, {\forall l \in \lset n \rset, } \sum_{l\in\integerbox{n}}{{ \p v}_l}\in \ber{n}{r}{m-1}\}.
\end{align}

The code family $\dber{n}{r}{m}\subseteq \Fb_2^{n^m}$, called the \emph{Dual Berman code with parameters $n, m$ and $r$}, is defined as
\begin{align*}
    &\dber{n}{ m}{m}\triangleq \Fb_2^{n^m}, &&\\
    &\dber{n}{ 0}{m}\triangleq \{(c, \hdots, c)\in \Fb_2^{n^m}\colon c\in \Fb_2\},
\end{align*} 
%%%%
and for $m\geq 2$ and $1\leq r\leq m-1,$ 
\begin{align}
    \nonumber\dber{n}{r}{m}\triangleq \{  ({ {\p u}}+{ {\p u}}_0|{ {\p u}}+{ {\p u}}_1|\hdots|{ {\p u}}+{ {\p u}}_{n-2}|{ {\p u}}) \colon \!  { {\p u}}_l\in &\dber{n}{r-1}{m-1}, {\forall l \in \lset n-1 \rset, }\\  
    \label{DualBermanRecDef}
    &~~~~~~~~~~~~~~~~~~~{ {\p u}}\in \dber{n}{r}{m-1}\}.
\end{align}
Note that if $n=2$, then we recover the $\RM$ codes from the above definitions. %\mathsf{DB}_2(r, m) = \RM(r, m)$ and $\ber{2}{r}{m} = \RM(m-r-1, m)$. Thus, the class of codes $\dber{n}{r}{m}$ includes the Reed-Muller codes. Also, by definition, the codes $\dber{n}{r}{m}$ and $\ber{n}{r}{m}$ are linear. 

\subsubsection{Properties of Berman and Dual Berman codes}
\label{PropertiesofBermananddualBerman}
We now present some properties of Berman and Dual Berman codes, which were proved in \cite{DBLakPra}. 
For, $1\leq r\leq m, $ we have the following containment properties:  $\ber{n}{r}{m}\subseteq \ber{n}{r-1}{m}$ and $\dber{n}{r-1}{m}\subseteq \dber{n}{r}{m}$. Further, $\ber{n}{r}{m}^\perp=\dber{n}{r}{m}$. Also, the code dimensions are as follows:  $\dim(\ber{n}{r}{m})=\sum_{w=r+1}^{m}\binom{m}{w}(n-1)^w,$ and $
\dim(\dber{n}{r}{m})=\sum_{w=0}^{r}\binom{m}{w}(n-1)^w$.  Their minimum distances are given as:  $d_{\min}(\ber{n}{r}{m})  = 2^{r+1}, ~~ 0\leq r \leq m-1,$ $d_{\min}(\dber{n}{r}{m}) = n^{m-r}, ~~0\leq r \leq m.$ Further, the Berman and Dual Berman codes are transitive codes. 

Let $H = \integerbox{n} = \{0, 1, \dots, n-1\}$. We then identify the $n^m$ coordinates of an arbitrary vector ${ \p v}\in \Fb_2^{n^m}$ using the $m$-tuples in $H^m$, i.e., ${ \p v}=(v_{ {\p i}}: { {\p i}}\in H^m)$. We also write $ \p v$ as a concatenation of $n$ vectors from, $\Fb_2^{n^{m-1}}, $ denoted by ${ \p v}=({ \p v}_0|\hdots|{ \p v}_{n-1})$. This concatenation representation can also be used recursively. % The subvector ${ \p v}_l\in \Fb_2^{n^{m-1}}$ is then identified recursively as follows. 
% \begin{itemize}
% \item For any $ {\p i'}\in H^{m-1}, $ the component of $ \p v_l$ indexed by $ {\p i'}$ is identified as $v_{l, { {\p i'}}}=v_{({ {\p i'}| l)}}$ which is the component of ${ \p v}$ indexed by $({\p i'}\, |\, l) \in H^m$. 
% \end{itemize} 
% Observe that the concatenation representation can be used recursively, for instance the subvector ${ \p v}_l\in\Fb_2^{n^{m-1}}$ can be written as a concatenation of $n$ subvectors from $\Fb_2^{n^{m-2}}$, and so on.  

%Already these are part of notation: 
%Let us denote the support of a vector $ {\p i} \in H^m$ as $\supp( {\p i}) = \{ k \in \lset m \rset :  {\p i}_k \neq 0\}$.  The weight or the Hamming weight of $ {\p i}$ is $\wt( {\p i}) = |\supp( {\p i})|$. 
We now present the bases for these codes. Towards that end, a partial ordering is defined among the vectors in $H^m$ as follows. For $\p j,\p i \in H^m$, we say that $\p j\preceq \p i$ (or equivalently, $\p i\succeq \p j$) if the components of $\p j$ in its non-zero locations are identical to the components of $\p i$ at the same locations. In other words,  we say $\p j\preceq \p i$ if the following is true.
%For $ {\p i}, {\p j} \in H^m$ we will say that `$ {\p i}$ contains $ {\p j}$', or equivalently, `$ {\p j}$ is contained in $ {\p i}$' if
\begin{equation*}
\supp ({\p j}) \subseteq \supp({\p i} ) \text{ and } {\p j}_{\supp ({\p j})} = {\p i}_{\supp (\p j)}.
\end{equation*} 
%That is, $ {\p i}$ contains $ {\p j}$ if for each $k \in \lset m \rset$ we have $j_k = 0$ or $j_k = i_k$.  
%We will denote this relation as $ {\p i} \succeq {\p j}$ or $ {\p j} \preceq {\p i}$. 
%Note that for any, $ {\p i} \in H^m$ there are exactly $2^{\wt( {\p i})}$ vectors contained by $ {\p i}$, and exactly $n^{m-\wt( {\p i})}$ vectors that contain $ {\p i}$. 

For $m\geq 1$ and some ${ {\p i'}}\in H^m$, the vector ${\p c}_m({ {\p i'}})\in \Fb_2^{n^m}$ is defined as the binary vector with components as follows. 
\begin{equation}
\label{cm}
c_m({ {\p i'}})_{ {\p i}}=\begin{cases}
1, & \text{if } {\p i} \preceq {\p i'}, \\
0, & \text{otherwise}.
\end{cases} ~~~~~~~~\forall { {\p i}}\in H^m. 
\end{equation}
Thus, ${\p c}_m({ {\p i'}})$ is the binary vector with support set $\{ {\p i} \in H^m:  {\p i} \preceq {\p i'} \}$. Further, note that $\wt({\p c}_m({ {\p i'}}))=2^{\wt(\p i')}$.  For any, $ {\p i'} \in H^m$ the vector $\p{d}_m( {\p i'}) \in \Fb_2^{n^m}$ with components is defined as follows. 
\begin{equation}
\label{dm}
d_m({ {\p i'}})_{ {\p i}}=\begin{cases}
1, &\text{if } {\p i} \succeq {\p i'} \\
0, &\text{otherwise}. 
\end{cases}
\end{equation}
Thus, $\p{d}_m( {\p i'})$ is the $n^{m-\wt(\p i')}$-weight vector with support $\{ {\p i} \in H^m:  {\p i} \succeq {\p i'}\}$. We are now ready to recall the definition of the bases of Berman codes. For $m\geq 1$, and $0\leq r\leq m-1$, the set 
%%%%%
$$\basisber{n}{r}{m} = \{ {\p c}_m({ {\p i'}})\colon \forall { {\p i'}}\in H^m ~\text{s.t.}~  r+1 \leq \wt({\p i'}) \leq m \}$$
%%%%
is the basis of the Berman Code $\ber{n}{r}{m}$ as shown in \cite{DBLakPra}. Similarly, from \cite{DBLakPra}, we have the following basis for $\dber{n}{r}{m}$.
$$\basisdber{n}{r}{m} = \left\{ \p{d}_m( {\p i'}):  \forall {\p i'} \in H^m ~\text{s.t.}~ 0 \leq \wt( {\p i'}) \leq r \right\}.$$
%%%%
%%%%%%
\section{Star products of Berman family of codes}
\label{sec: starprodBermanCodes}
We now show the star products of different combinations of Berman and Dual Berman codes. These will be used in Section \ref{sec: tPIRusingBerandDBer} to obtain new $t$-PIR schemes. We first present the result for the star product of two Dual Berman codes.
%%%%
\begin{lemma}
\label{lemma: dualstarpro}
Let $n\geq 2$ be any integer. Let $r_1$ and $r_2$ be two non-negative integers such that $r_1 + r_2 \leq m$. Then 
\begin{align*}
    \dber{n}{r_1}{m} \star \dber{n}{r_2}{m} = \dber{n}{r_1+r_2}{m}.
\end{align*}
\end{lemma}

\begin{IEEEproof}
To prove the Lemma, it is sufficient to prove the following two claims. 
\begin{claim}
\label{claimDB1}
\textbf{$\dber{n}{r_1}{m} \star \dber{n}{r_2}{m} \subseteq \dber{n}{r_1+r_2}{m}$}.
\end{claim}

\begin{claim}
\label{claimDB2}
Every basis vector in $\basisdber{n}{r_1+r_2}{m}$ can be  obtained as a star product of appropriate basis vectors in $\basisdber{n}{r_1}{m}$ and $\basisdber{n}{r_2}{m}$. In other words, for any ${\p j}\in H^m$ such that $w_H({\p j}) \leq r_1+ r_2$, there exists tuples ${\p j}_1,{\p j}_2\in H^m$ with $w_H({\p j}_1) \leq r_1 $ and $w_H({\p j}_2) \leq r_2 $, such that 
%%%%
\begin{align}
\label{eqn: starproductbasesconditionDber}
{\p d_m}({\p j}) = {\p d_m}({\p j}_1) \star {\p d_m}({\p j}_2).
\end{align}
%%%%
\end{claim}

We now shall prove Claim \ref{claimDB1} by using induction on $m$. Let $m = 1$, the only possible choices for $r_1, r_2$ are 
(a) $r_1 = r_2 = 0,$ or (b) $\{r_1, r_2\} = \{0, 1\}.$ As $\dber{n}{ 0}{1}$ is just the repetition code of length $n$, in both these cases, we get 
$\dber{n}{ 0}{1} \star \dber{n}{ 0}{1} \subseteq \dber{n}{ 0}{1}$, and $\dber{n}{ 0}{1} \star \dber{n}{ 1}{1} \subseteq \dber{n}{ 1}{1}.$ Therefore, the claim is valid for the base case. 

We now assume that the claim is valid till $m$, for all values of $r_1, r_2$ such that $0\leq r_1, r_2 \leq \;r_1 +r_2 \leq m$. We need to show that the claim holds for $m+1$ also, for all non-negative values $r_1,r_2$ such that $r_1+r_2\leq m+1$. Now, if $r_i=0$ for any $i\in\{1,2\}$, then as $\dber{n}{0}{m+1}$ is the repetition code, the claim is easily seen to hold. Thus, we assume $r_1,r_2>0$. Further, observe that, in this case, if $r_1+r_2=m+1$, then the statement of Claim \ref{claimDB1} holds trivially, as the code $\dber{n}{r_1+r_2}{m+1}$ is the entire space $\Fb^{n^{m+1}}$. Hence, we assume $r_1+r_2\leq m$, enabling us to apply the induction hypothesis effectively.  

Consider two arbitrary codewords ${\p c}_1 \in \dber{n}{r_1}{m + 1}$ and ${\p c}_2 \in \dber{n}{r_2}{m + 1}$. By the recursive definition in (\ref{DualBermanRecDef}), we can write  ${\p c}_1= ( {\p u} + {\p u}_0| {\p u} + {\p u}_1| \hdots\hspace{0.2cm}| {\p u}+ {\p u}_{n-2}| {\p u}),$ and ${\p c}_2= ( \p v + \p v_0| \p v + \p v_1| \hdots\hspace{0.2cm}| \p v + \p v_{n-2}| \p v),$ where ${\p u}_i\in \dber{n}{r_1-1}{m}$ and $ \p v_i\in \dber{n}{r_2-1}{m}$, $\forall i$, while ${\p u}\in \dber{n}{r_1}{m}$ and ${\p v}\in \dber{n}{r_2}{m}$. Then the star product of these two codewords is: 
% \pk{See the formatting of the expression below and use of hdots. Use this as much as you can}
\begin{align*}
 {\p c}_1 \star {\p c}_2  & = (( {\p u} + {\p u}_0) \star( \p v + \p v_0)|( {\p u} + {\p u}_1)\star( \p v + \p v_1)| \hdots\hspace{0.2cm} | ( {\p u} + {\p u}_{n-2})\star( \p v + \p v_{n-2}) | {\p u} \star \p v). &&
\end{align*}

Note that $( {\p u} + {\p u}_i) \star( \p v + \p v_i) = {\p u} \star \p v + {\p u}_i \star \p v + {\p u} \star \p v_i + {\p u}_i \star \p v_i.$ Assign $ {\p x} = {\p u} \star \p v$  and  $ {\p x}_i = {\p u}_i \star \p v + {\p u} \star \p v_i + {\p u}_i \star \p v_i, \; \forall i \in \integerbox{n-2}.$ We can thus write $
 {\p c}_1 \star {\p c}_2  =( {\p x} + {\p x}_0| {\p x} + {\p x}_1| .  .  .  | {\p x} + {\p x}_{n-2}| {\p x}).$ 
 
 Observe that, by our induction hypothesis and invoking the recursive definition of the Dual Berman codes in (\ref{DualBermanRecDef}), we have that $ \p x = {\p u} \star \p v \in \dber{n}{r_1+r_2}{m}\text{, and } {\p x_i} = {\p u}_i \star \p v + {\p u} \star \p v_i + {\p u}_i \star \p v_i \in \dber{n}{r_1+r_2 - 1}{m}, \forall i$.  Thus, by (\ref{DualBermanRecDef}), we can conclude that $ {\p c}_1 \star {\p c}_2 \in \dber{n}{r_1+r_2 }{m + 1}.$ Therefore, $\dber{n}{r_1}{m + 1} \star \dber{n}{r_2}{m + 1} \subseteq \dber{n}{r_1+r_2 }{m + 1}$. Now Claim \ref{claimDB1} is proved.

We now prove Claim \ref{claimDB2}. Without loss of generality, assume $\max(r_1, r_2) = r_1$. We are given a basis vector ${\p d_m}(\p j)$ of $\dber{n}{r_1+r_2}{m}$ and want to find corresponding basis vectors ${\p d_m}(\p j_1)$ and ${\p d_m}(\p j_2)$ with $\wt(\p j_i)\leq r_i$ for $i\in\{1,2\}$, such that (\ref{eqn: starproductbasesconditionDber}) is satisfied. 

If $w_H({\p j}) \leq r_1$ 
then we can take ${\p j}_1 = {\p j}$ and ${\p j}_2 = \textbf{0}$. Observe that in this case, we have ${\p d_m}({\p j}_1)  \in \basisdber{n}{r_1}{m}$ and ${\p d_m}({\p j}_2)  \in \basisdber{n}{r_2}{m}$. As ${\p d_m}({\p j}_2)$ is the all-one codeword in this case, we see that (\ref{eqn: starproductbasesconditionDber}) is satisfied. 

Now consider that $r_1 + 1 \leq w_H( {\p j}) \leq r_1 + r_2$. Then we can write $ {\p j}$ as $ {\p j} = {\p j}_1 + {\p j}_2 $  where $ {\p j}_1, {\p j}_2 \in H^m$, $\supp( {\p j}_1) \cap \supp( {\p j}_2) = \emptyset$, $w_H( {\p j}_1) = r_1 \; \text{and} \; w_H( {\p j}_2) = w_H({\p j}) - r_1\leq r_2$.  As $w_H( {\p j}_1) = r_1$ we have that ${\p d_m}( {\p j}_1) \in \basisdber{n}{r_1}{m}$. Similarly, as $w_H( {\p j}_2) \leq r_2$, we have ${\p d_m}( {\p j}_2) \in \basisdber{n}{r_2}{m}$. Let us denote  ${\p d_m}( {\p j}_1, {\p j}_2)\triangleq {\p d_m}( {\p j}_1) \star {\p d_m}( {\p j}_2)$. We will show that ${\p d_m}( {\p j}_1, {\p j}_2)={\p d_m}(\p j)$. 

Now, the support  $\supp({\p d_m}( {\p j}_1, {\p j}_2))$ will be those locations in which both ${\p d_m}( {\p j}_2) \; \text{and} \; {\p d_m}( {\p j}_1)$ are both 1. Thus, such locations $ {\p i} \in H^m$ should satisfy $ {\p i}\succeq {\p j}_1 $ and $ {\p i} \succeq {\p j}_2$. 

% \pk{note that now, the support is $\supp$ not $\supp$. Please use this only everywhere.} 

Since the $\supp( {\p j}_1) \cap \supp( {\p j}_2) = \emptyset$ we have that
\begin{align*}
\{ {\p i} \in H^m:  {\p i}\succeq {\p j}_1, \; {\p i} \succeq {\p j}_2\} = \{ {\p i} \in H^m:  {\p i}\succeq {\p j}_1 + {\p j}_2 = {\p j} \}.
\end{align*}

Hence, we have for all, $\p i\in H^m$
\begin{align*}
    {d_m}( {\p j}_1, {\p j}_2)_ { {\p i}} =  
    \begin{cases}
        1, \; {\p i} \succeq {\p j}, \\
        0, \;\text{otherwise.}
    \end{cases}
\end{align*}
which is exactly the same as the definition of ${\p d_m}( {\p j})$ by equation (\ref{dm}). This proves Claim \ref{claimDB2}. This concludes the proof of the lemma.
% \pk{Lemma \ref{lemma: dualstarpro} is now checked with proof. Note that the label is changed to {lemma: dualstarpro} everywhere. Do similar things everywhere.}
\end{IEEEproof}
%%%%
\begin{remark}
For $n = 2$, $\dber{n}{r}{m}=\RM(r, m)$. Hence, we recover the result from \cite{tPIRIEEE}, that
$\RM (r_1, m) \star \RM(r_2, m) = \RM(r_1 + r_2, m).$
\end{remark}
%%%%%%%
Lemma \ref{lemma: bermandualbermanstarpro} shows the star product of a Berman code with a Dual Berman code, in a specific parameter regime that we use for obtaining one of our PIR schemes.

\begin{lemma}
\label{lemma: bermandualbermanstarpro}
Let $n\geq 2$. Let $r_1$ and $r_2$ be two non-negative integers such that $0 \leq r_2 \leq r_1  \leq m$. Then 
\begin{align*}
    \ber{n}{r_1}{m} \star \dber{n}{r_2}{m} = \ber{n}{r_1-r_2}{m}.
\end{align*}
\end{lemma}
\begin{IEEEproof}
Clearly, the Lemma is true if $r_1=m$, as $\ber{n}{m}{m}$ is, just the zero-dimensional subspace. Hence, we assume throughout that $r_1\leq m-1$. As with the proof of Lemma \ref{lemma: dualstarpro}, we prove Lemma \ref{lemma: bermandualbermanstarpro} by proving the following two claims. 
\begin{claim}
\label{BDB1}
$\ber{n}{r_1}{m} \star \dber{n}{r_2}{m} \subseteq \ber{n}{r_1-r_2}{m}.$
\end{claim}
\begin{claim}
\label{BDB2}
Every basis vector in $\basisber{n}{r_1-r_2}{m}$ can be  obtained as a linear combination of the star products of the appropriate basis vectors of $\basisber{n}{r_1}{m}$ and $\basisdber{n}{r_2}{m}$.
\end{claim}

We shall now prove the claim \ref{BDB1} by using induction on $m$. 

Base case:  Let $m = 1$ and the possible value for $r_1, \; r_2$ are $r_1 = 0, r_2 = 0$.  We can observe that
\begin{align*}
    \ber{n}{0}{1} * \dber{n}{ 0}{1} \subseteq \ber{n}{0}{1}.
\end{align*}
This is true as $\ber{n}{0}{1}$ is a single parity check code of length $n$, while $\dber{n}{ 0}{1}$ is a repetition code of length $n$. Therefore, the base case with $m=1$ is true. 

Assume that the claim is valid until $m$, for any $r_1$ and $r_2$ such that $0 \leq r_2 \leq r_1  \leq m-1 $. We have to prove the claim for $m+1$, for all $r_1,r_2$ such that $0\leq r_2\leq r_1\leq m$ (recall that the $r_1=m+1$ scenario is trivial). 

Let $ {\p u} \in \ber{n}{r_1}{m+1}, \; \p v \in \dber{n}{r_2}{m + 1}$ be two arbitrary codewords from the respective codes. We know that from the recursive definition of $\ber{n}{r}{m}$ and $\dber{n}{r}{m}$ in (\ref{BermanRecDef}) and (\ref{DualBermanRecDef}) that
%%%
\begin{align*}
 {\p u} &= ( {\p b}_0, {\p b}_1, \hdots, {\p b}_{n-1}), ~\text{and} \\
 \p v &= ( {\p c} +{\p c}_0, {\p c} + {\p c}_1, \hdots, {\p c} + {\p c}_{n-2}, {\p c})
\end{align*} 
%%%%
where $\p b_i, \p c, \text{ and } \p c_i $ are defined as per (\ref{BermanRecDef}) and (\ref{DualBermanRecDef}). Thus, we have 
\begin{align*}{\p u} \star \p v &=  ( {\p b}_0 \star ( {\p c} + {\p c}_0),\hdots, {\p b}_{n-2} \star ( {\p c} + {\p c}_{n-2}), {\p b}_{n-1} \star {\p c}).
\end{align*}

From the induction hypothesis, we see that ${\p b}_{i} * {\p c} \in \ber{n}{r_1-r_2-1}{m}, \forall i\in\lset n-1 \rset$. We can also observe that ${\p b}_i \star ( {\p c} + {\p c}_i) = {\p b}_i \star {\p c} + {\p b}_i \star {\p c}_i$ where ${\p b}_i \star {\p c} \in \ber{n}{r_1-r_2-1}{m}$ and ${\p b}_i \star {\p c}_i \in \ber{n}{r_1-r_2}{m}$, $\forall i \in \lset n-1 \rset$, again from the induction hypothesis. We know from the containment property in Subsection \ref{PropertiesofBermananddualBerman}, that $\ber{n}{r_1-r_2}{m} \subseteq \ber{n}{r_1-r_2-1}{m}$. Hence, ${\p b}_i \star {\p c} + {\p b}_i \star {\p c}_i \in \ber{n}{r_1-r_2-1}{m}$, $\forall i \in \lset n-1 \rset$. 

Thus, to show, $ {\p u} \star \p v \in \ber{n}{r_1-r_2}{m+1}$ we only need to show the following. 
\begin{align}
\label{Eq:  u star v}
    & \sum_{i = 0}^{n-2} {\p b}_i \star ( {\p c} + {\p c}_i) + {\p b}_{n-1} * {\p c} \in \ber{n}{r_1-r_2}{m}.&& 
\end{align}

Now, the sum $\sum_{i = 0}^{n-2} {\p b}_i \star ( {\p c} + {\p c}_i) + {\p b}_{n-1} * {\p c} =  \sum_{i = 0}^{n-1} {\p b}_i \star {\p c} + \sum_{i = 0}^{n-2} {\p b}_i \star {\p c}_i$. Since we have by the induction hypothesis that
${\p b}_i \star {\p c}_i \in \ber{n}{r_1-r_2}{m},$ we thus have that $ \sum_{i = 0}^{n-2} {\p b}_i \star {\p c}_i \in \ber{n}{r_1-r_2}{m}$. We can further simplify $\sum_{i = 0}^{n-1} {\p b}_i \star {\p c} =  (\sum_{i = 0}^{n-1} {\p b}_i) \star {\p c}$. From the definition of codeword $ {\p u}$, we know that $\sum_{i = 0}^{n-1} {\p b}_i  \in \ber{n}{r_1}{m}$. We can now conclude that 
\begin{align*}
     \sum_{i = 0}^{n-1} {\p b}_i \star {\p c} \in \ber{n}{r_1-r_2}{m}
\end{align*} from the induction hypothesis. 
Thus, (\ref{Eq:  u star v}) holds. Hence, ${\p u} \star \p v \in \ber{n}{r_1-r_2}{m+1}$. 
Therefore, $\ber{n}{r_1}{m+1} \star \dber{n}{r_2}{m + 1} \subseteq \ber{n}{r_1-r_2}{m+1}$. This concludes the proof of claim \ref{BDB1}.

%\pk{3/2:  Srikar, please read this part of the proof of Claim \ref{BDB2} upto the $r=1$ case carefully. I made some considerable changes though I retained the spirit of your argument.}

We now prove claim \ref{BDB2}. Clearly, if $r_2=0$, Claim \ref{BDB2} is true, since the code $\dber{n}{0}{m}$ is just the repetition code which has the singular basis vector $(1,\hdots,1)$. %Similarly, it is easy to see that Claim \ref{BDB2} is true when $r_1=m$. Hence, we assume $1\leq r_2\leq r_1 < m$.
Hence, we assume $1\leq r_2$. 

First, we look at the special case of $r_2 = 1$. We seek to generate the basis vectors of $\ber{n}{r_1-1}{m}$ by linear combinations of the star products of the basis vectors of $\ber{n}{r_1}{m}$ and $\dber{n}{1}{m}$. Consider a basis vector ${\p c_m}( {\p j}' ) \in  \basisber{n}{r_1-1}{m}$. Note that, if  $\wt({\p j'})\geq r_1+1$, then by definition ${\p c_m}( {\p j}' )$ is a basis vector of $\ber{n}{r_1}{m}$ also. Note that the all-one codeword is the basis vector $\p d_m(\p 0)$  of $\dber{n}{1}{m}$. Hence, we can obtain the basis vector ${\p c_m}( {\p j}' ) \in  \basisber{n}{r_1-1}{m}$ as the star product of the basis vector  ${\p c_m}( {\p j}' ) \in  \basisber{n}{r_1}{m}$ and the basis vector $\p d_m(\p 0)$  of $\dber{n}{1}{m}$, in the case of $\wt({\p j'})\geq r_1+1$. 

Now looking at the case when $\wt({\p j'})=r_1.$ In this case, we can write $\p j'=\p j-\p k$, for some $\p j$ and $\p k$  such that we have $ \wt(\p j)=r_1+1\leq m$ and $\wt(\p k) = 1$  and 
\begin{align}
\label{conditiononK}
    \supp( {\p k} ) \subseteq \supp( {\p j} ), \; ~~\text{and} \; j_i = k_i, \; \text{~for~} i \in \supp(\p k). 
\end{align}

%Consider any ${\p c_m}( {\p j} ) \in  \basisber{n}{r_1}{m} $. Let ${\p d_m}( {\p k} ) \in \basisdber{n}{1}{m}$ 
Note that $\p c_m(\p j)\in \basisber{n}{r_1}{m}$ and $\p d_m(\p k)\in \basisdber{n}{1}{m}.$ We then see that the following is true. 
 \begin{align}
 \nonumber
     {\p c_m}( {\p j} ) \star {\p d_m}( {\p k} ) &= {\p c_m}( {\p j} ) + {\p c_m}( {\p j}-{\p k} )\\
 \label{theoremeq1}
 &={\p c_m}( {\p j} )\star {\p d_m}( {\p 0} )  + {\p c_m}(\p j').
\end{align}
Rearranging  (\ref{theoremeq1}), we thus get 
\begin{align}
\label{eqn: lemmaberdber}
{\p c_m}(\p j')={\p c_m}( {\p j} ) \star {\p d_m}( {\p k} ) +{\p c_m}( {\p j} )\star {\p d_m}( {\p 0} ).
\end{align}
By prior arguments, the R.H.S of (\ref{eqn: lemmaberdber}) is clearly a linear combination of star products of basis vectors from $\ber{n}{r_1}{m}$ and $\dber{n}{1}{m}$. This proves Claim \ref{BDB2}, and hence Lemma \ref{lemma: bermandualbermanstarpro} too, for the case of $r_2=1$. Now, using Lemma \ref{lemma: dualstarpro}, the $r_2=1$ case we just proved, and by the associativity property of the star product, we get the following for the case of $r_2\geq 2$. 
 \begin{align*}
    &\ber{n}{r_1}{m} \star \dber{n}{r_2}{m} &&\\
    &= \ber{n}{r_1}{m} \star ( \dber{n}{ 1}{m} \star \dber{n}{r_2-1}{m})&& \\
    & =(\ber{n}{r_1}{m} \star \dber{n}{ 1}{m} ) \star \dber{n}{r_2-1}{m} &&\\
    & = \ber{n}{r_1-1}{m} \star \dber{n}{r_2-1}{m}. &&
 \end{align*}
By using similar arguments repeatedly, we get
 \begin{align*}
   &\ber{n}{r_1}{m} \star \dber{n}{r_2}{m} &&\\
   & = \ber{n}{r_1-r_2}{m}\star \dber{n}{ 0}{m} &&\\
   & = \ber{n}{r_1-r_2}{m}, &&
\end{align*}
where the last equality holds as $\dber{n}{0}{m}$ is just the repetition code. This concludes the proof for claim \ref{BDB2}. 

\end{IEEEproof}
\begin{remark}
     For $n = 2$, $\ber{2}{r_1}{m}=\RM(r_1, m)^{\bot}=\RM(m-r_1-1,m)$ and $\dber{2}{r_2}{m}=\RM(r_2, m)$.  Thus, we see that the star product of the two codes from Lemma \ref{lemma: bermandualbermanstarpro} is, $\ber{2}{r_1-r_2}{m}$ which is $\RM(m-(r_1-r_2)-1,m)$. This matches with the known result for star products of Reed-Muller codes, from \cite{tPIRIEEE}.
\end{remark}

Lemma \ref{thm: otherstarproductsnotuseful1}, \& \ref{thm: otherstarproductsnotuseful2}, identifies the star products of Berman code combinations not considered so far (observe that for $n=2$ ($\RM$ codes), Lemma \ref{lemma: dualstarpro} essentially covers all cases). These are not useful for PIR, as they do not result in a non-zero rate PIR scheme. Still, we present these results for the sake of completeness.

\begin{lemma}
\label{thm: otherstarproductsnotuseful1}
    Let $ n \geq 3$, and $r_1, r_2$ be two non-negative numbers such that $0\leq r_1, r_2 \leq  m$, then $\ber{n}{r_1}{m} \star \ber{n}{r_2}{m} = \Fb_2^{n^m}.$
\end{lemma}
\begin{IEEEproof}
The result is trivial if $r_1$ or $r_2$ is equal to $m$, as the Berman code $\ber{n}{m}{m}$ code is just the zero-dimensional subspace. Hence, we focus on the case when $r_1,r_2\leq m-1$ and $m\geq 1$. 

We shall prove the theorem by obtaining the standard basis of $\Fb_2^{n^m}$ from the star product of the basis vectors of $\ber{n}{r_1}{m} \; \text{and} \; \ber{n}{r_2}{m}$. We know from the containment properties of the Berman codes given in Subsection \ref{PropertiesofBermananddualBerman} that $\ber{n}{r}{m}\subseteq \ber{n}{r-1}{m}$.  Given $0\leq r_1, r_2 \leq m-1$, if we can show that the standard basis of $\Fb_2^{n^m}$ can be derived from the star product of the basis vectors of  $\ber{n}{m-1}{m} \; \text{and} \; \ber{n}{m-1}{m}$, it will be true for any value of $0\leq r_1, r_2 \leq m-1$. We denote the standard basis vectors of $\Fb_2^{n^m}$ as $\{{\p e}_{\p j}\colon \forall {\p j} \in H^m\}$. Let us now consider the star product $\ber{n}{m-1}{m} \star \ber{n}{m-1}{m}$, where any basis vector ${\p c_m}( {\p j} ) \in  \basisber{n}{m-1}{m} $ is such that $\wt({\p j}) = m$.  We shall show the standard basis of $\Fb_2^{n^m}$ can be derived from the star product of the basis vectors of  $\ber{n}{m-1}{m} \; \text{and} \; \ber{n}{m-1}{m}$ in multiple steps. 

\textbf{Step $0$: }
     
     To get ${\p e}_{\p 0}$ we can take ${\p c_m}( {\p j}) \in  \basisber{n}{m-1}{m} $ and ${\p c_m}( {\p l}) \in  \basisber{n}{m-1}{m}$, such that $l_i \neq j_i, \forall i$. Clearly, such choices for $\p l$ and $\p j$ exist, since $n\geq 3$. Then, by (\ref{cm}), we see that
    \begin{align*}
        {\p c_m}( {\p j}) \star {\p c_m}( {\p l}) =  {\p e}_{\p 0}.
    \end{align*}
 
\textbf{step $k$ (for each $k\in\{1,\hdots,m\}$): }

    In this step, we obtain all ${\p e}_{ \p v}, \; \forall \p v\in H^m $  such that $w_H(\p v) = k$ as linear combinations of the star products of the basis vectors of $\ber{n}{m-1}{m}$, assuming that the Step $k-1$ is already completed. To obtain ${\p e}_{ \p v}$ for some $\p v\in H^m$ with $w_H(\p v) = k$, we take some arbitrary ${\p c_m}( {\p j}), \; {\p c_m}( {\p l})\in  \basisber{n}{m-1}{m}$, such that 
    $l_i=j_i=v_i, \forall i\in\supp(\p v)$, and $l_i\neq j_i$ in the remaining $m-k$ locations.  Once again, such choices for $\p l$ and $\p j$ exist, by definition of $\ber{n}{m-1}{m}$, and because $n\geq 3$. We then see that,
    \begin{align*}
        {\p c_m}( {\p j}) \star {\p c_m}( {\p l}) = {\p c_m}(\p v) \in \Fb_2^{n^m}.
    \end{align*}  By (\ref{cm}), we see that the vector ${\p c_m}(\p v)$ is just the sum of the vectors in $\{{\p e}_{\p y}: \forall {\p y}\in H^m~\text{s.t.}~{\p y}\preceq{\p v}\}$. Except for $\p e_{\p v}$, all the other elements in this set are already obtained in the previous Steps (steps $0$ to $(k-1)$). Hence, we will be able to obtain ${\p e}_{ \p v}$ as linear combinations of the star products of the basis vectors of $\ber{n}{m-1}{m}$. This concludes the proof.
\end{IEEEproof}

\begin{lemma}
\label{thm: otherstarproductsnotuseful2}
    Let $n\geq 2$, and $r_1, r_2,m$ be non-negative numbers such that $0\leq r_1 < r_2 \leq m$, then $\ber{n}{r_1}{m} \star \dber{n}{r_2}{m} = \Fb_2^{n^m}.$
\end{lemma}

\begin{IEEEproof}
Clearly, $r_2\geq 1$.  As with Lemma \ref{thm: otherstarproductsnotuseful1}, we shall prove this lemma by deriving the standard basis of $ \Fb_2^{n^m}$ from the star product of the basis vectors of  $\ber{n}{r_1}{m} \; \text{and} \; \dber{n}{r_2}{m}$. We know from the properties of the Berman codes given in Subsection \ref{PropertiesofBermananddualBerman} that $\ber{n}{r_2-1}{m}\subseteq \ber{n}{r_1}{m}$.  Thus, if we can show that the standard basis of $\Fb_2^{n^m}$ can be derived from the star product of the basis vectors of  $\ber{n}{r_2-1}{m} \; \text{and} \; \dber{n}{r_2}{m}$, we would have proved the Lemma. This is precisely what we now do.

Let us denote the standard basis vectors of $\Fb_2^{n^m}$ as $\{{\p e}_{\p j}\colon \forall {\p j} \in H^m\}$.  We shall show that each vector in this standard basis set can be derived from the star product of the basis vectors of  $\ber{n}{r_2-1}{m} \; \text{and} \; \dber{n}{r_2}{m}$ in multiple steps.

\textbf{Step $0$: } In this step, we obtain the $\binom{m}{r_2}(n-1)^{r_2}$ standard basis vectors $\{{\p e}_{\p j}:  \forall {\p j}\in H^m \text{~s.t.~} \wt({\p j})=r_2\}$.  To do this, observe that for any $\p j\in H^m$ such that  $\wt({\p j})=r_2$, we have ${\p c_m}(\p j)\in \basisber{n}{r_2 - 1}{m}$ and ${\p d_m}(\p j)\in \basisdber{n}{r_2}{m}.$ further, note that
    \begin{align*}
        {\p c_m}(\p j) \star {\p d_m}(\p j) = {\p e}_{\p j}, 
    \end{align*}
    which follows by (\ref{cm}) and (\ref{dm}). This completes Step $0$. 
    
\textbf{step $k$ (for each $k\in\{1,\hdots,m\}$): }

In this step, we obtain $\binom{m}{r_2 - k}(n-1)^{r_2 - k}$ standard basis vectors, $\{{\p e}_{\p j'}:  \forall \p j' \text{~s.t.~}\wt(\p j') = r_2-k\},$ assuming that we have already completed steps up to $k-1$.

To do this, observe that for any $ \p j'$ with $\wt(\p j') = r_2-k$, there exists some $\p j$ such that $\wt(\p j)=r_2$ such that $\p j'\preceq\p j$. Clearly, we see that ${\p c_m}(\p j)\in \basisber{n}{r_2 - 1}{m}$. Further, ${\p d_m}(\p j')\in \basisdber{n}{r_2}{m}.$ Let $\p x={\p c_m}(\p j) \star {\p d_m}(\p j')$. 

Now, from the structure of the vectors ${\p c_m}(\p j)$ and ${\p d_m}(\p j')$, observe that $\p x$ can be written as the sum of the standard basis vectors $\{{\p e}_{\p v}\colon \forall \p j' \preceq \p v \preceq \p j\}$. In this collection, all the standard basis vectors except $\p e_{\p j'}$ have been obtained in the prior steps (as they have weights in $\{r_2-k+1,\hdots,r_2\}$). Thus, from $\p x$ and the standard basis vectors in $\{{\p e}_{\p v}\colon \forall \p j' \preceq \p v \preceq \p j\}\setminus \{\p e_{\p j'}\}$, we can obtain ${\p e}_{\p j'}$ as a linear combination of the star products' basis vectors of $\dber{n}{r_2}{m}$ and $\ber{n}{r_2-1}{m}$.

Note that if $r_2=m$, the proof is complete. Thus, the next two steps are to be executed only if $r_2<m$.

\textbf{Step $r_2 + 1$: }

In this step, we obtain the $\binom{m}{r_2 +1}(n-1)^{r_2 + 1}$ standard basis vectors $\{{\p e}_{\p j}:  \forall {\p j}\in H^m \text{~s.t.~} \wt({\p j})=r_2 + 1\}$.  To do this, observe that for any $\p j\in H^m$ such that  $\wt({\p j})=r_2 + 1$, there exists some $\p j'$ such that $\wt(\p j') =  r_2$, and $\p j'\preceq\p j$, where ${\p c_m}(\p j)\in \basisber{n}{r_2 - 1}{m}$ and ${\p d_m}(\p j')\in \basisdber{n}{r_2}{m}.$ Then, it is easy to see that
\begin{align*}
    {\p c_m}(\p j) \star {\p d_m}(\p j') = {\p e}_{\p j} + {\p e}_{\p j'}. 
\end{align*}
Since we obtained the standard basis vectors ${\p e}_{\p j'}$ in Step $0$ (as $ \wt(\p j') = r_2$), we can obtain ${\p e}_{\p j}$ as the linear combination ${\p c_m}(\p j) \star {\p d_m}(\p j')+{\p e}_{\p j'}$.

\textbf{Step $r_2 + l$ (for each $l$ from $2$ to $m- r_2 $) : }

In this step, we obtain the $\binom{m}{r_2 + l}(n-1)^{r_2 + l}$ standard basis vectors $\{{\p e}_{\p j}:  \forall {\p j}\in H^m \text{~s.t.~} \wt({\p j})=r_2 + l\}$. To do this, observe that for any $\p j\in H^m$ such that  $\wt({\p j})=r_2 + l$, there exists some $\p j'$ such that $\wt(\p j') =  r_2$, and $\p j'\preceq\p j$, where ${\p c_m}(\p j)\in \basisber{n}{r_2 - 1}{m}$ and ${\p d_m}(\p j')\in \basisdber{n}{r_2}{m}.$  Let   $\p x= {\p c_m}(\p j) \star {\p d_m}(\p j').$ Observe, by the structure of vectors ${\p c_m}(\p j)$ and ${\p d_m}(\p j')$ that $\p x$ is just the sum of the vectors $\{{\p e}_{\p v}\colon \forall \p j' \preceq \p v \preceq \p j\}$. Except for the vector $\p e_{\p j}$ in this collection, all the other standard basis vectors have already been obtained in the previous steps as linear combinations of the star products of the basis vectors of the two concerned codes. Using these and $\p x$, we can see therefore that $\p e_{\p j}$ can also be obtained as a linear combination of the star products' basis vectors of $\dber{n}{r_2}{m}$ and $\ber{n}{r_2-1}{m}$. This completes the proof.
\end{IEEEproof}

\begin{remark}
     For $n = 2$, $\ber{2}{r_1}{m}$ is $\RM(r_1, m)^{\bot}$ and ${\cal C}_2(r_2, m)$ is $\RM(r_2, m)$ where $r_2 > r_1$. Applying the star product to these two linear codes, we get,
\begin{align*}
    \ber{2}{r_1}{m} \star {{\cal C}}_2(r_2, m)  = \RM(m - r_1 - 1, m)  \star \RM(r_2, m) &= \RM(m - r_1 - 1 +r_2, m) &&\\
    & = \RM( m + (r_2 - r_1) - 1, m) &&
\end{align*}
As $r_2 > r_1$, $r_2 - r_1 \geq 1$, $\ber{2}{r_1}{m} \star {{\cal C}}_2(r_2, m) = \RM(m, m) =\Fb_2^{2^m}$, which aligns with the statement of our theorem
\end{remark}

\section{New $t$ - private information retrieval via star products of Berman and Dual Berman codes}
\label{sec: tPIRusingBerandDBer}
In this section, we present the main results of this paper, which involve applying the results involving the star products of Berman and Dual Berman codes to the $t$-PIR framework developed in \cite{tPIRIEEE}.
\begin{table*}[htbp]
\centering
\begin{tabular}{||c|c|c|c|c||} 
 \hline
& & & & \\ 
Storage Code & Retrieval Code & $\rst$ & $\rpir$ & Collusion parameter:  $t$ \\
& & & & \\
\hline\hline
& & & & \\
$ \dber{n}{r_{\cal C}}{m}$ & $ \dber{n}{ r_{\cal D}}{m} $ & $\dfrac{\sum_{i = 0}^{r_{\cal C}} \binom{m}{i} (n-1)^i}{n^m}$ & $\dfrac{\sum_{i = r_{\cal C} + r_{\cal D} +1}^{m} \binom{m}{i} (n-1)^i}{n^m}$ & 
 $2^{r_{\cal D} + 1} - 1 $  \\
& & & & \\
\hline
& & & & \\
$ \dber{n}{r_{\cal C}}{m}$ & $ \ber{n}{r_{\cal D}}{m} $   & $\dfrac{\sum_{i = 0}^{r_{\cal C}} \binom{m}{i} (n-1)^i}{n^m}$     & $\dfrac{\sum_{i = 0}^{r_{\cal D} - r_{\cal C}} \binom{m}{i} (n-1)^i}{n^m}$    & $ n^{m - r_{\cal D}} - 1 $ \\
& ($r_{\cal D} \geq r_{\cal C}$)& & & \\
\hline
& & & & \\
$ \ber{n}{r_{\cal C}}{m} $ & $ \dber{n}{r_{\cal D}}{m} $  & $\dfrac{\sum_{i = r_{\cal C} + 1}^{m} \binom{m}{i} (n-1)^i}{n^m}$ & $\dfrac{\sum_{i = 0}^{r_{\cal C} - r_{\cal D}} \binom{m}{i} (n-1)^i}{n^m}$    & $2^{r_{\cal D} + 1} - 1$ \\ 
&($r_{\cal C} \geq r_{\cal D}$) & & & \\
\hline
\end{tabular}
\caption{\small PIR parameters $\rst$, $\rpir$, and $t$ for possible choices of Storage ($\cal{C}$) and Retrieval ($\cal{D}$) codes from class of Berman Codes.}
\label{table: Maintheoremtable}
\hrule
\end{table*}

\begin{theorem}
For any integers $n\geq 2$ and $m\geq 1$, there exists star product based $t$-PIR schemes on $n^m$ servers, enabling privacy from $t$-server collusion, with storage-retrieval code pairs and parameters $(\rst, \rpir, t)$ as shown in each of the three rows of Table \ref{table: Maintheoremtable}. 
\end{theorem}

\begin{IEEEproof}
We give the proof for the scheme presented in the first row of Table \ref{table: Maintheoremtable}. The proofs corresponding to the other rows of Table \ref{table: Maintheoremtable} follow similarly, by using Lemma \ref{lemma: bermandualbermanstarpro} in conjunction with Theorem \ref{Theorem:  Starproduct}.

For ${\cal C}= \dber{n}{r_{\cal C}}{m}$ being the storage code, and ${\cal D} = \dber{n}{ r_{\cal D}}{m}$ being the retrieval code, we have ${\cal C} \star {\cal D} = \dber{n}{r_{\cal C} + r_{\cal D}}{m}$, by Lemma \ref{lemma: dualstarpro}. Since ${\cal C}$ and ${\cal C} \star {\cal D}$ are both in the class of Berman codes, they are transitive, as shown in \cite{DBLakPra}. Hence, by Theorem \ref{Theorem:  Starproduct} given in Section \ref{sec:  Preliminaries}, there exists a star product PIR scheme with storage-retrieval code pair as the chosen codes $\cal C$ and $\cal D$, with PIR rate as follows.
\begin{align*}
    \rpir = \frac{\dim (({\cal C} \star {\cal D})^{\perp})}{n^m} &= \frac{\dim \; (\ber{n}{r_{\cal C}+r_{\cal D}}{m})}{n^m}&&\\
    &= \frac{\sum_{i = r_{\cal C} + r_{\cal D} + 1}^{m} \binom{m}{i} (n-1)^i}{n^m}.&&
\end{align*}
Further, the storage rate is obtained as 
\begin{align*}
    \rst = \frac{\dim (\cal C)}{n^m}& = \frac{\dim \; (\dber{n}{r_{\cal C}}{m}) }{n^m}=\frac{\sum_{i = 0}^{r_{\cal C}} \binom{m}{i} (n-1)^i}{n^m}.&&
\end{align*}
Finally, the collusion parameter $t$ is calculated as $d_{\min}({{\cal D}^{\perp}}) - 1 = d_{\min}({\dber{n}{ r_{\cal D}}{m}^{\perp}}) - 1  = \left(2^{r_{\cal D} + 1} - 1\right).$
\end{IEEEproof}

%\enlargethispage{-2cm}
\subsection{Numerical comparisons}
We now present some numerical examples of the three new PIR schemes shown in Table \ref{table: Maintheoremtable}, in Tables \ref{table: berdber}, \ref{table: dberdber},  and \ref{table: dberber}. In these tables, the top row (from the second column onwards) runs through some choices for the pair of parameters $(n,m)$, while the first column runs through some choices for the pair of parameters $r_{\cal C},r_{\cal D}\in\{0,\hdots,m\}$. The entry inside the table corresponding to a row defined by the pair $(r_{\cal C},r_{\cal D})$ and a column given by the pair $(n,m)$ denotes the triple $(t,\rst,\rpir)$ (number of colluding servers tolerated, the storage rate, and the PIR rate) of the star product scheme defined by the respective storage and retrieval code. Note that in all these tables, the quantity $n^m$ represents the number of servers, which is the same as the length of the storage and retrieval codes. The choices for $(n,m)$ are chosen to keep the number of servers $n^m$ as comparable as possible. When $n=2$, these codes are the $\RM(r,m)$ codes and the PIR schemes derived are those given in \cite{tPIRIEEE}, which we compare our schemes with. We observe that our Berman codes-based PIR schemes perform better in some cases, in one or more parameters. For instance, when the storage code ${\cal C}=\ber{n}{r_{\cal C}}{m}$ and  retrieval code ${\cal D}=\dber{n}{r_{\cal D}}{m}$ (Table \ref{table: berdber}), the PIR rate $\rpir$ is higher for the case when $(r_{\cal C}, r_{\cal D}) = (0,0)$ and $(n,m) = (3,3)$ (or $(5,2)$)  when compared to the RM codes. Similarly, as observed from Table \ref{table: dberdber}, we can improve the $\rst$ of various schemes obtained via different $({r_{\cal C}}, {r_{\cal D}})$ pairs, while keeping the $\rpir$ similar to that of Reed-Muller codes and for the same $t$ parameter. We also see  in Table \ref{table: dberber} that our schemes achieve higher $\rpir$ and $\rst$ compared to the Reed-Muller case across various pairs of $({r_{\cal C}}, {r_{\cal D}})$, trading it off with smaller values of the colluding parameter $t$.
%%%

\section{Discussion}
\label{sec: discussion}
In this work, we have presented a new class of PIR schemes based on the framework developed in \cite{FriejHollantiSIAM,tPIRIEEE}, using Berman codes and showcased their advantages. For this purpose, we have characterized the star products of pairs of codes from the class of Berman codes. The schemes we present have flexibility in the number of servers, the PIR rate, the storage rate, and the collusion parameter $t$. A possible future work  is to investigate other retrieval schemes for the PIR schemes for Berman-coded storage systems. Further, it would also be interesting to extend our schemes to the Byzantine and unresponsive servers' scenario, similar to what was done for $\RM$ codes in \cite{PIRRMIEEE}. Further enhancements are also possible, such as via applying the refine and lift technique as shown in \cite{OneSHot} to our schemes.
\clearpage

\begin{table*}[htbp]
\centering
\begin{tabular}{||c|c|c|c|c||} 
 \hline
 $(r_{\cal C}, r_{\cal D})$ & $(n,m) = (2,5)$ & $(n,m) = (3,3)$ & $(n,m) = (5,2)$  & $(n,m) = (6,2)$\\ [0.5ex]
  & (RM codes of length $32$, \cite{tPIRIEEE}) & & & \\ [0.5ex]
 \hline\hline
(0,0) & (1, 0.969, 0.031) & (1, 0.963, 0.037) & (1, 0.96, 0.04) & (1, 0.972, 0.028) \\ 
\hline
(1,0) & (1, 0.812, 0.188) & (1, 0.741, 0.259) & (1, 0.64, 0.36) & (1, 0.694, 0.306) \\
\hline
(1,1) & (3, 0.812, 0.031) & (3, 0.741, 0.037) &  (3, 0.64, 0.04) & (3, 0.694, 0.028) \\
% \hline
% (2,0) & (1, 0.5, 0.5) & (1, 0.296, 0.704) & - & -\\
 \hline
\end{tabular}
\caption{Comparison of achievable triples $(t,\rst,\rpir)$ for the case of storage code ${\cal C}=\ber{n}{r_{\cal C}}{m}$ and ${\cal D}=\dber{n}{r_{\cal D}}{m}$. }
\hrule
\label{table: berdber}
\end{table*}

\begin{table*}[htbp]
\centering 
\begin{tabular}{||c|c|c|c|c||} 
 \hline
 $(r_{\cal C}, r_{\cal D})$ & $(n,m) = (2,5)$ & $(n,m) = (3,3)$ & $(n,m) = (5,2)$  & $(n,m) = (6,2)$\\ [0.5ex]
  & (RM codes of length $32$, \cite{tPIRIEEE}) & & & \\ [0.5ex]
 \hline\hline
(0,0) & (1, 0.03, 0.96) & (1, 0.03, 0.96) & (1, 0.04, 0.96) & (1, 0.02, 0.97) \\ 
\hline
(0,1) & (3, 0.03, 0.81) & (3, 0.03, 0.74) & (3, 0.04, 0.64) & (3, 0.02, 0.69)\\
\hline
% (0,2) & (7, 0.03, 0.5) & (7, 0.03, 0.29) & (7, 0.04, 0.0) & (7, 0.02, 0.0)\\
\hline
(1,0) & (1, 0.18, 0.81) & (1, 0.25, 0.74) & (1, 0.36, 0.64) & (1, 0.3, 0.69)\\
% \hline
% (1,1) & (3, 0.18, 0.5) & (3, 0.25, 0.29) & (3, 0.36, 0.0) & (3, 0.3, 0.0)\\
% \hline
% (2,0) & (1, 0.5, 0.5) & (1, 0.70, 0.29) & (1, 1.0, 0.0) & (1, 1.0, 0.0) \\
 \hline
\end{tabular}
\caption{Comparison of achievable triples $(t,\rst,\rpir)$ for the case of storage code ${\cal C}=\dber{n}{r_{\cal C}}{m}$ and ${\cal D}=\dber{n}{r_{\cal D}}{m}$. }
\hrule
\label{table: dberdber}
\end{table*}

\begin{table*}[htbp]
\centering
\begin{tabular}{||c|c|c|c|c||} 
 \hline
  $(r_{\cal C}, r_{\cal D})$ & $(n,m) = (2,5)$ & $(n,m) = (3,3)$ & $(n,m) = (5,2)$  & $(n,m) = (6,2)$\\ [0.5ex]
  & (RM codes of length $32$, \cite{tPIRIEEE}) & & & \\ [0.5ex]
 \hline\hline
(0,0) & (31, 0.031, 0.031) & (26, 0.037,0.037) & (24, 0.04, 0.04) & (35, 0.028, 0.028)\\
\hline
(0,1) & (15, 0.031, 0.188) & (8, 0.037, 0.259) & (4, 0.04, 0.36) & (5, 0.028, 0.306) \\
% \hline
% (0,2) & (7, 0.031, 0.5) & (2, 0.037, 0.704) & (0, 0.04, 1.0) & (0, 0.028, 1.0) \\
\hline
(1,1) & (15, 0.188, 0.031) & (8, 0.259, 0.037)& (4, 0.36, 0.04) & (5, 0.306, 0.028) \\
\hline
\end{tabular}
\caption{Comparison of achievable triples $(t,\rst,\rpir)$ for the case of storage code ${\cal C}=\dber{n}{r_{\cal C}}{m}$ and ${\cal D}=\ber{n}{r_{\cal D}}{m}$. }
\hrule
\label{table: dberber}
\end{table*}

\clearpage
\bibliographystyle{IEEEtran}
\bibliography {ArxivISIT23}
\end{document}